\documentclass[conference, final]{IEEEtran}

\usepackage{scalerel}
\usepackage{tikz}
\usetikzlibrary{svg.path}

\definecolor{orcidlogocol}{HTML}{A6CE39}
\tikzset{
  orcidlogo/.pic={
    \fill[orcidlogocol] svg{M256,128c0,70.7-57.3,128-128,128C57.3,256,0,198.7,0,128C0,57.3,57.3,0,128,0C198.7,0,256,57.3,256,128z};
    \fill[white] svg{M86.3,186.2H70.9V79.1h15.4v48.4V186.2z}
                 svg{M108.9,79.1h41.6c39.6,0,57,28.3,57,53.6c0,27.5-21.5,53.6-56.8,53.6h-41.8V79.1z M124.3,172.4h24.5c34.9,0,42.9-26.5,42.9-39.7c0-21.5-13.7-39.7-43.7-39.7h-23.7V172.4z}
                 svg{M88.7,56.8c0,5.5-4.5,10.1-10.1,10.1c-5.6,0-10.1-4.6-10.1-10.1c0-5.6,4.5-10.1,10.1-10.1C84.2,46.7,88.7,51.3,88.7,56.8z};
  }
}

\newcommand\orcidicon[1]{\href{https://orcid.org/#1}{\mbox{\scalerel*{
    \begin{tikzpicture}[yscale=-1,transform shape]
        \pic{orcidlogo};
    \end{tikzpicture}
}{|}}}}

\usepackage{cite}
\usepackage{amsmath,amssymb,amsfonts}
\usepackage{algorithmic}
\usepackage{graphicx}
\usepackage{textcomp}
\usepackage{ctable}
\usepackage{xcolor}
\usepackage{subfig}

\usepackage{hyperref}

\DeclareRobustCommand*{\IEEEauthorrefmark}[1]{%
  \raisebox{0pt}[0pt][0pt]{\textsuperscript{\footnotesize\ensuremath{#1}}}}

\def\BibTeX{{\rm B\kern-.05em{\sc i\kern-.025em b}\kern-.08em
    T\kern-.1667em\lower.7ex\hbox{E}\kern-.125emX}}
\newcolumntype{M}[1]{>{\centering\arraybackslash}m{#1}}

\begin{document}

\title{Virtual SAR: A Synthetic Dataset for Deep Learning based Speckle Noise Reduction Algorithms}

\author{

\IEEEauthorblockN{Shrey Dabhi\IEEEauthorrefmark{1}\IEEEauthorrefmark{2} \orcidicon{0000-0002-4364-9892}\,, Kartavya Soni\IEEEauthorrefmark{1} \orcidicon{0000-0002-9553-4835}\,, Utkarsh Patel\IEEEauthorrefmark{1} \orcidicon{0000-0001-7402-6092}\,, Priyanka Sharma\IEEEauthorrefmark{1} \orcidicon{0000-0002-1934-5805}\,, Manojkumar Parmar\IEEEauthorrefmark{2}\IEEEauthorrefmark{3} \orcidicon{0000-0002-1183-4399}}

\IEEEauthorblockA{\IEEEauthorrefmark{1}Department of Computer Science and Engineering\\
Institute of Technology, Nirma University\\
Ahmedabad, India\\
\{16bit039, 16bit087, 16bit083, priyanka.sharma\}@nirmauni.ac.in}

\IEEEauthorblockA{\IEEEauthorrefmark{2}Engineering Technology Strategy (RBEI/ETS)\\
Robert Bosch Engineering and Business Solutions Private Limited,\\
Bengaluru, India\\
manojkumar.parmar@bosch.com}

\IEEEauthorblockA{\IEEEauthorrefmark{3}HEC Paris,\\
Jouy-en-Josas, France\\
manojkumar.parmar@hec.edu}

\\[-6.0ex]

}

\maketitle

\begin{abstract}
Synthetic Aperture Radar (SAR) images contain a huge amount of information, however, the number of practical use-cases is limited due to the presence of speckle noise in them. In recent years, deep learning based techniques have brought significant improvement in the domain of denoising and image restoration. However, further research has been hampered by the lack of availability of data suitable for training deep neural network based systems. With this paper, we propose a standard way of generating synthetic data for the training of speckle reduction algorithms and demonstrate a use-case to advance research in this domain.
\end{abstract}


\section{Introduction}

Synthetic aperture radar (SAR) is a form of radar that is used for creating 2D or 3D reconstructions of objects. It is generally used in remote sensing.  It relies on electromagnetic waves in the microwave spectrum to generate images and hence can see through obstructions. SAR sensor is mounted on a moving platform that travels in the time taken by the emitted pulse to return to the camera, thereby creating a larger perceived aperture. The resolution of the image relies on the aperture of the sensor, regardless of the nature of the aperture, so it can generate relatively higher resolution images with a smaller aperture. SAR's ability to produce high-quality images at night and in adverse weather conditions with relatively smaller physical aperture provides an advantage over traditional optical and infrared imaging systems.

SAR images inherently suffer from a multiplicative noise, called speckle. This type of noise is the result of constructive and destructive interference of coherent reflections scattered by small reflectors inside each resolution cell. The effect of speckle tends to weaken for very high-resolution systems since the number of elemental scatterers within a resolution cell decreases \cite{sar-tutorial-1}. The presence of speckle noise makes image processing and computer vision tasks relatively difficult. Hence, it is important to remove speckle from SAR images to improve the performance and efficiency of various computer vision tasks such as segmentation, object detection, classification, and recognition.

In further sections, we discuss our study and provide our conclusions. In Section 2 we elaborate our study of SAR images, nature of speckle noise and existing algorithms used for speckle reduction. We also share our motivation behind curating a new dataset for this task. In Section 3 we give the reasoning for our approach towards designing the dataset. In Section 4, we train one of the state-of-the-art deep neural network based systems using our dataset and provide the visual and qualitative results for the same.

\section{Literature Study}

The most commonly used model to understand distributed scatterers causing the speckle noise in SAR images is \cite{goodman1976some}:
\begin{equation}
    Y = N X,
\end{equation}
where $Y \in \mathbb{R}^{W \times H}$ is the observed image intensity, $X \in \mathbb{R}^{W \times H}$ is the noise-free image, and $N \in \mathbb{R}^{W \times H}$ is the normalized fading speckle noise random variable. Assuming that the SAR image is an average of $L$ looks, one common assumption on $N$ is that it follows a Gamma distribution with unit mean and variance $\frac{1}{L}$. It has the following probability distribution function \cite{ulaby2019handbook}
\begin{equation}
    p(N) = \left(\frac{N L}{e}\right)^{L} \frac{e^{-N}}{N} \frac{1}{\Gamma(L)}, N \geq 0, L \geq 1,
\end{equation}
where $\Gamma(\cdot)$ denotes the Gamma function.

Researchers have used statistical techniques like multilooking \cite{oliver2004understanding} \cite{Jakowatz1996}, pixel based filtering \cite{lee-filter} \cite{kuan-filter} \cite{frost-filter} \cite{ppb}, wavelet based filtering \cite{sarbm3d}, etc. for removing speckle. Some of these methods \cite{sar-cnn} transfer the image into the logarithmic domain to transform multiplicative noise into additive noise, but it has its limitations which have been discussed in the upcoming section. Due to the non-local nature of the processing being carried out by some of these methods \cite{lee-filter} \cite{kuan-filter} \cite{frost-filter} \cite{ppb}, they are not able to preserve sharp features and edges, making the task of the further processing even more difficult.

In recent times there has been a meteoric rise in the usage of deep learning in image restoration and enhancement tasks like denoising and super-resolution. Efforts on despeckling have also been able to beat the state-of-the-art in this domain \cite{sar-cnn} \cite{id-cnn}, yet the reproducibility and reliability are very low due to the lack of a publicly available standard dataset for this task.\footnote {Unlike image classification, image segmentation, video classification, object detection, etc. no standard dataset is available on which researchers can develop neural network based SAR image despeckling techniques.}

Moving and Stationary Target Acquisition and Recognition (MSTAR) \cite{mstar} system has been used by many researchers to improve the performance of Automatic Target Recognition (ATR) systems operating over SAR images \cite{mstar-use1} \cite{mstar-use2}, but the images from this model have not been used for training denoising models due to the lack to diversity in images. As described in \cite{BALZ2015102}, a small number of commercial and open-source real-time simulators for SAR images are also available. However, they require a lot of time and computational power for generating the diverse and huge dataset, required by deep learning based image denoising techniques. An attempt to create a systematic framework for the evaluation of any general speckle reduction algorithm is available in G. Di Martino et al. \cite{6351163}, but it still does not resolve the problem of the lack of a training set for deep learning based despeckling techniques.

Authors of Image Despeckling Convolutional Neural Network (ID-CNN) \cite{id-cnn} and Fractional ID-CNN (FID-CNN) \cite{fid-cnn} have used a combination of Uncompressed Colour Image Dataset (UCID) \cite{ucid}, Berkeley Segmentation Data Set and Benchmarks 500 (BSDS500) \cite{bsd-500} and Northwestern Polytechnical University REmote Sensing Image Scene Classification (NWPU-RESISC45) \cite{nwpu-resisc45}, after adding a static amount of speckle noise, for training their models. In F. Lattari et al. \cite{sar-unet}, they have used images from UC Merced Land Use Dataset \cite{uc-merced} and PatternNet \cite{patternnet} to create training and testing datasets respectively. Hence, it is evident that there is a lot of variation in the datasets used by each research study for training and testing their algorithms, which makes it very difficult for anyone to objectively compare the results and determine which is the best algorithm.

It is relatively easy to find actual SAR images from the National Aeronautics and Space Administration's (NASA) Alaska Satellite Facility (Vertex portal) and the European Space Agency's (ESA) Copernicus portal. They have published a huge amount of data from various missions like Sentinel - 1 (C band) and ALOS PALASAR (L band) under open access. They also provide higher resolution data from the missions for research and study requests.

Even though actual SAR images are so freely available in the public domain, it is practically impossible to use them directly for training purposes. This is because all the captured SAR images inherently contain speckle noise and it is not possible to extract a completely noise free patches from actual data and create pairs of noisy and clean image patches required for training a neural network. Converting SAR images to a format usable by the deep learning frameworks is a slow process and requires significant computational power.

Aerial images are closely related to SAR images in terms of structure and type of features that they contain. This makes them the next best candidate for generating simulated SAR images. REmote Sensing Image Scene Classification (RESISC), created by Northwestern Polytechnical University (NWPU) \cite{nwpu-resisc45} is one of the largest collections of aerial images, containing images of 45 different classes with 700 images belonging to each class.

After a very comprehensive study, we are of opinion that research on deep neural networks in this domain is in a nascent state. We feel the need to establish an initial standard dataset which can be used by all the researchers to advance the field, because:
\begin{itemize}
    \item SAR images are difficult to handle due to their size.
    \item SAR images cannot be directly used for training deep neural networks due to a lack of pairs of clean and noisy images.
    \item Currently, there is no standard reference dataset available for this task.
\end{itemize}
We try to address all these issues by developing the very first reference dataset, for this image processing task, as described in the next section.

\section{Proposed Dataset}

We have created a dataset using all the 31500 images from NWPU-RESISC45 \cite{nwpu-resisc45}. A. Moreira et al. \cite{sar-tutorial-1} have established that the variance of the speckle noise model depends upon the actual value of the signal, i.e. the value of noise and the amplitude of the original signal are correlated. F. Argenti et al. \cite{sar-tutorial-2} discuss the following alternative model used for representing speckle noise:
\begin{equation}
    J = \eta I
\end{equation}
\begin{equation}
    J = I + (\eta - 1) I
\end{equation}
\begin{equation}
    J = I + K
\end{equation}
where $J$ is the observed image intensity, $I$ is the noise free image and $K = (\eta - 1)I$ accounts for the multiplicative speckle component of the SAR image and $\eta$ is randomly sampled from a uniform distribution with mean $\mu$ and variance $\sigma$. Furthermore, S. Abramov et al. \cite{Abramov14} suggests that $\sigma \in [0.55, 0.9]$.

We propose to divide images from each category of NWPU-RESISC45 dataset \cite{nwpu-resisc45} in a way that images of each category have noise belonging to the entire possible spectrum of values of $\sigma$ in the training set to make sure that each level of noise has enough diversity. We include the cases of lower noise as compared to the actual SAR images to induce a moderate amount of regularization while training the neural network. It helps the network to become robust while avoiding over-fitting the noisy images.

We are of the opinion that 37 images from publicly available USC-SIPI Image Database \cite{usc-sipi}, Volume 2: Aerials can be borrowed as the cross-validation set. The variance $\sigma$ can be set to any value, but we choose to use the default value provided by Matlab's implementation of $\texttt{imnoise()}$ function \cite{matlab:2018}.

\section{Experimental Results}

\begin{figure}[ht]
    \captionsetup{justification=centering}
    \centering
    \includegraphics[width=0.5\textwidth]{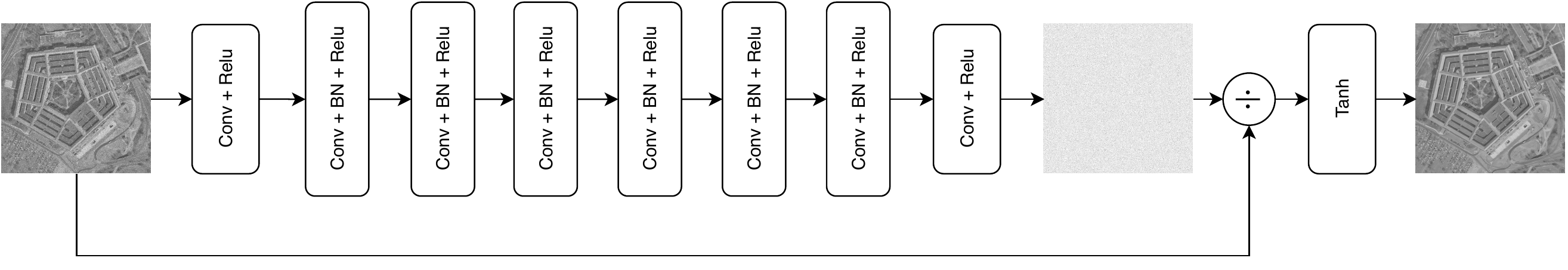}
    \caption{Architecture of ID-CNN}
    \label{fig:idcnn}
\end{figure}

In this section, we provide details of the experiments we carried out using our synthetically generated dataset. We decided to baseline our dataset using a CNN architecture inspired by the current state-of-the-art, ID-CNN \cite{id-cnn}. A possible solution to the multiplicative nature of the noise would be to transfer the image to the logarithmic domain. However, that would introduce the problem of negative infinity values where-ever the value of a pixel is exactly $0$. Therefore, we decide to skip this step and try to learn a direct mapping from the original image to the residual noise and thereby to the clean image. P. Wang et al. \cite{id-cnn} and K. Zhang et al. \cite{7839189} provide more detailed reasoning for the design of the architecture. The discussion of the same is beyond the scope of this paper. We dropped the regularizing total variation (TV) term from the loss function as the clean images in the dataset already regularizing the training process. Fig.~\ref{fig:idcnn} shows the CNN architecture we borrowed from ID-CNN. The CNN was trained using ADAM optimizer with the default learning rate $0.001$ and mean squared error as the loss function. The model was trained on a private Kaggle kernel, which provides a CPU with 2 cores and an Nvidia Tesla P100 with 16 GB of RAM.

We compare the performance of the CNN with that of the following 5 despeckling algorithms: Lee filter \cite{lee-filter}, Kuan filter \cite{kuan-filter}, Frost filter \cite{frost-filter}, SAR-BM3D \cite{sarbm3d} and PPB \cite{ppb}. Note that all the parameters are set as suggested in their corresponding papers.

\begin{figure*}[ht]
    \captionsetup{justification=centering}
    \centering
    \subfloat[b][Groundtruth]{\includegraphics[width=0.24\textwidth]{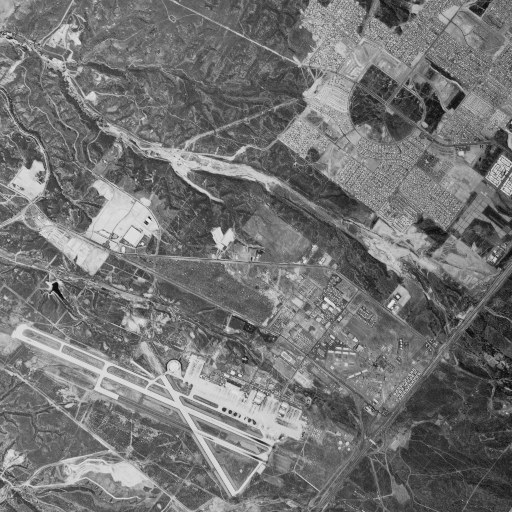}}
    \hspace{0pt}
    \subfloat[b][Noisy Image]{\includegraphics[width=0.24\textwidth]{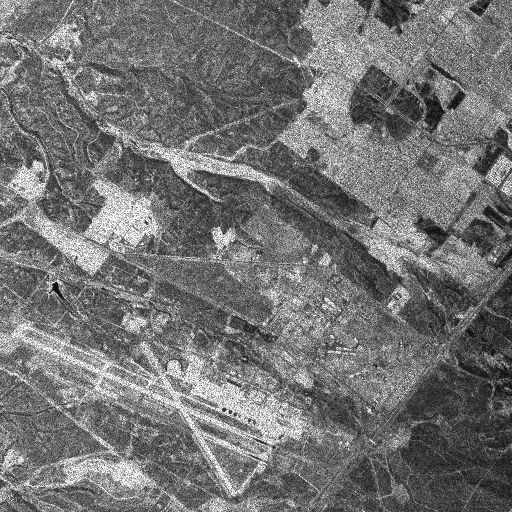}}
    \hspace{0pt}
    \subfloat[b][ID-CNN]{\includegraphics[width=0.24\textwidth]{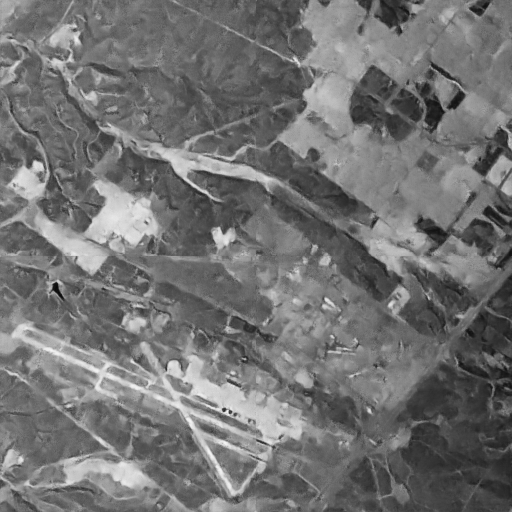}}
    \hspace{0pt}
    \subfloat[b][SAR-BM3D]{\includegraphics[width=0.24\textwidth]{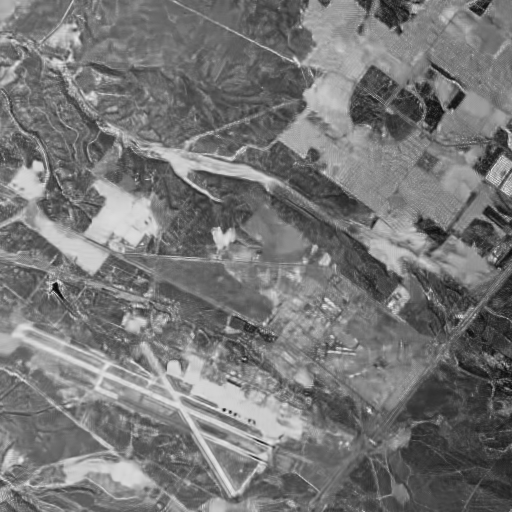}}
    \hspace{0pt}
    \subfloat[b][Lee Filter]{\includegraphics[width=0.24\textwidth]{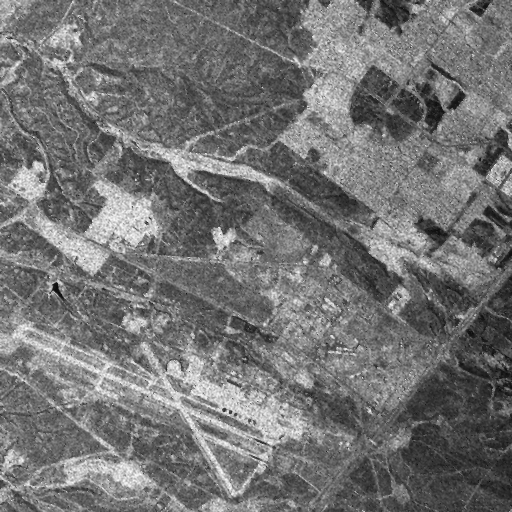}}
    \hspace{0pt}
    \subfloat[b][Kuan Filter]{\includegraphics[width=0.24\textwidth]{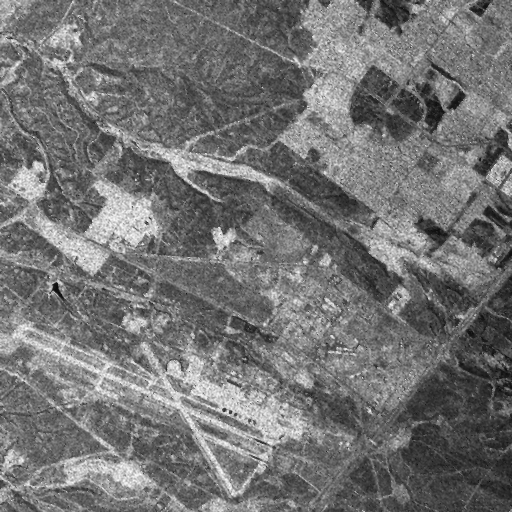}}
    \hspace{0pt}
    \subfloat[b][Frost Filter]{\includegraphics[width=0.24\textwidth]{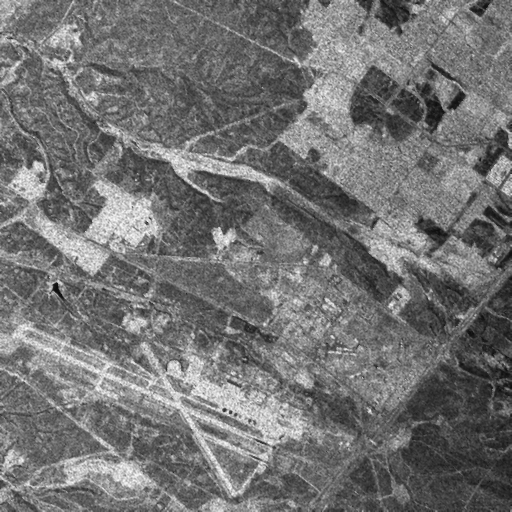}}
    \hspace{0pt}
    \subfloat[b][PPB Filter]{\includegraphics[width=0.24\textwidth]{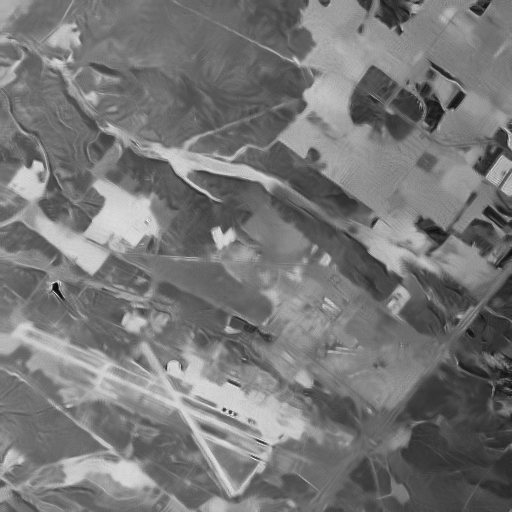}}
    \caption{Visual results of despeckling techniques on an image from USC SIPI Image Database \cite{usc-sipi}}
    \label{fig:simar_outputs}
\end{figure*}

Fig.~\ref{fig:simar_outputs} and Table.~\ref{tab:metrics} contain the qualitative and quantitative results of our experiment respectively. The functions $\texttt{peak\char`_signal\char`_noise\char`_ratio()}$ and $\texttt{structural\char`_similarity()}$ from \textbf{scikit-image} package are used for measuring Peak Signal to Noise Ratio (PSNR) and Structural Similarity Index (SSIM) respectively. Equivalent Number of Looks (ENL) is a metric to compare the relative quality of the image after denoising in the absence of a ground-truth image. It is a parameter of multilook SAR images, which describes the degree of averaging applied to the SAR measurements during data formation and sometimes also post-processing. The following formula has been used for calculating ENL of the images \cite{oliver2004understanding}:
\begin{equation}
    ENL = \frac{(mean)^{2}}{variance}
\end{equation}

The biggest motivation for using deep neural networks is to reduce the pre-processing time required for SAR image based pipelines. The time taken by all the methods to denoise a $1024 \times 1024$ image is given in Table~\ref{tab:metrics}. The experiments to measure the time taken for denoising were carried out on a high-performance mobile workstation running Windows 10 operating system. It is equipped with 32 GB of random access memory and Intel\textsuperscript{\textregistered} Core\textsuperscript{TM} i7-8650U CPU @ 2.11 GHz with 4 physical cores and 8 logical cores. We used Python 3.7.6 compiled for MSC v.1916 for a 64-bit processor and Matlab 2019b with Image Processing Toolbox.


\begin{table*}[ht]
    \captionsetup{justification=centering}
    \caption{Quantitative results of despeckling techniques on an image from USC SIPI Image Database \cite{usc-sipi}}\label{tab:metrics}
    \begin{tabular*}{\textwidth}{M{0.1\textwidth} | M{0.1\textwidth} | M{0.1\textwidth} | M{0.1\textwidth} | M{0.1\textwidth} | M{0.1\textwidth} | M{0.1\textwidth} | M{0.1\textwidth}}
    \specialrule{.2em}{.1em}{.1em}
        Metric & Noisy & Lee filter & Kuan filter & Frost filter & PPB & SAR-BM3D & CNN \\ \specialrule{.1em}{.05em}{.05em}
        ENL & 11.237 & 20.852 & 20.950 & 21.080 & 30.351 & \textbf{34.990} & 31.362 \\
        PSNR & 18.182 & 22.965 & 22.989 & 22.945 & 25.830 & 26.025 & \textbf{27.119} \\
        SSIM & 0.294 & 0.457 & 0.458 & 0.453 & 0.594 & 0.617 & \textbf{0.661} \\
        Time taken & - & 38.37 s & 57.89 s & 79.65 s & 137.28 s & 159.74 s & \textbf{3.22 s} \\ \specialrule{.2em}{.1em}{.1em}
    \end{tabular*}
\end{table*}

\section{Conclusion}

Synthetic Aperture Radar (SAR) images have innumerable applications but cannot be used to their fullest till date due to a lack of better denoising techniques. We provide a novel synthetic dataset to accelerate the development of better deep neural networks for denoising of SAR images. We also demonstrate the usage of the synthetic data and improvement in the performance of the state-of-the-art convolutional neural network based speckle reduction system. It can potentially change the scope of work and research in this domain.

\section*{Conflict of Interest}

The authors declare no conflict of interest.

\section*{Acknowledgments}

We want to thank Prof. Anitha Modi from Insitute of Technology, Nirma University, India for her valuable comments, contributions and continued support to the project. We are would like to thank the National Aeronautics and Space Administration (NASA) and the European Space Agency (ESA) for providing the data from the Sentinel-1 and the ALOS PALSAR missions and the SNAP 7.0 software. We are grateful to all experts for providing us with their valuable insights and informed opinions ensuring completeness of our study.

\bibliographystyle{IEEEtran}
\bibliography{IEEEabrv,references}

\appendix

The actual dataset for public release purpose is in the process of deployment at cloud services. Until the please write to \texttt{16bit039@nirmauni.ac.in} for access to the dataset.

\end{document}